\documentclass[amssymb,amsmath,twocolumn,prl]{revtex4}

\usepackage{color,graphicx,bm}
\usepackage{overpic}

\newcommand{\ua}{\uparrow}
\newcommand{\da}{\downarrow}

\begin{document}

\title{
Pairing Theory of Striped Superconductivity\\[0mm]
}

\author{{}
Florian Loder, Siegfried Graser, Arno P. Kampf, and Thilo Kopp
\vspace{0,5cm}}

\affiliation{Center for Electronic Correlations and Magnetism,\\ 
Institute of Physics, University of Augsburg,
D-86135 Augsburg, Germany}

\date{\today}

\begin{abstract}
Striped high-$T_{\rm c}$ superconductors such as  La$_{2-y-x}$Nd$_{y}$Sr$_x$CuO$_4$ and La$_{2-x}$Ba$_{x}$CuO$_4$ near $x=1/8$ show a fascinating competition between spin and charge order and superconductivity. A theory for these systems therefore has to capture both the spin correlations of an antiferromagnet and the pair correlations of a superconductor. For this purpose we present here an effective Hartree-Fock theory incorporating both electron pairing with finite center-of-mass momentum and antiferromagnetism. We show that this theory reproduces the key experimental features such as the formation of the antiferromagnetic stripe patterns at $7/8$ band filling or the quasi one-dimensional electronic structure observed by photoemission spectroscopy.
\end{abstract}


\maketitle

Unidirectional charge- and spin-density modulations were predicted
\cite{ZaanenHF} for doped transition metal oxides even before their
experimental discovery in layered nickelates \cite{TranquadaNickel} and the
rare-earth doped cuprate La$_{2-x}$Sr$_x$CuO$_4$ \cite{TranquadaLNdSCO} and 
eventually in La$_{2-x}$Ba$_x$CuO$_4$~\cite{Fujita04}. Stripe patterns emerge 
as a compromise between correlation driven antiferromagnetism and 
an optimized kinetic energy gain for mobile charge carriers \cite{KivelsonRMP}. 
Charge and spin stripe textures were indeed obtained in various approximate 
model analyses of correlated electron systems, but it has remained unresolved 
which model systems sustain stable groundstate solutions with stripes and 
superconductivity. Here we report a pairing theory for the coexistence of 
charge and spin stripes
with $d$-wave superconductivity that results from an extension
of the BCS theory of superconductivity with an attractive pairing interaction
for tight-binding electrons moving on a square lattice. Charge and spin 
densities and the local pairing amplitudes adjust spatially in a stripe pattern
with transverse sign change 
for the antiferromagnetic (AF) order parameter. Hopping 
anisotropy weakens or even destroys superconductivity, as observed in the 
low-temperature tetragonal phase of cuprate superconductors~\cite{Buchner94,Tranquada97,Tranquada08,Hucker10}.

Transport experiments in the high-temperature superconductor 
La$_{2-x}$Ba$_x$CuO$_4$ for $x=1/8$ uncovered a sequence of thermal phase 
transitions \cite{LiPRL,Tranquada08,Hucker10}. Charge- and spin-stripe order
emerges sequentially upon cooling before two dimensional (2D) superconducting (SC) fluctuations set in 
which ultimately lead to 3D superconductivity below 4K. These measurements 
provided compelling evidence for what has since been called a striped 
superconductor. The subsequently developed theory for the striped 
superconductor introduced the concept of a pair density wave (PDW) in which the 
order parameter for the pairing of electrons in a superconductor is spatially 
modulated with respect to the center of mass coordinate of the electron pair
\cite{Berg09,Berg:2009}. This implies that Cooper pairs with finite momenta 
$\pm{\bf q}$ form accompanied by a charge-density modulation with wavenumber
$2{\bf q}$ \cite{Loder:2010}. The phenomenological characteristics of a pair 
density wave state with unidirectional charge modulation were either explored 
with respect to symmetry aspects and the nature of defects 
\cite{Berg:2009,Agterberg08} or its spectral properties \cite{Baruch:2008}. 

\begin{figure*}[]
\centering
\vspace{2.5mm}
\begin{overpic}
[width=0.64\columnwidth]{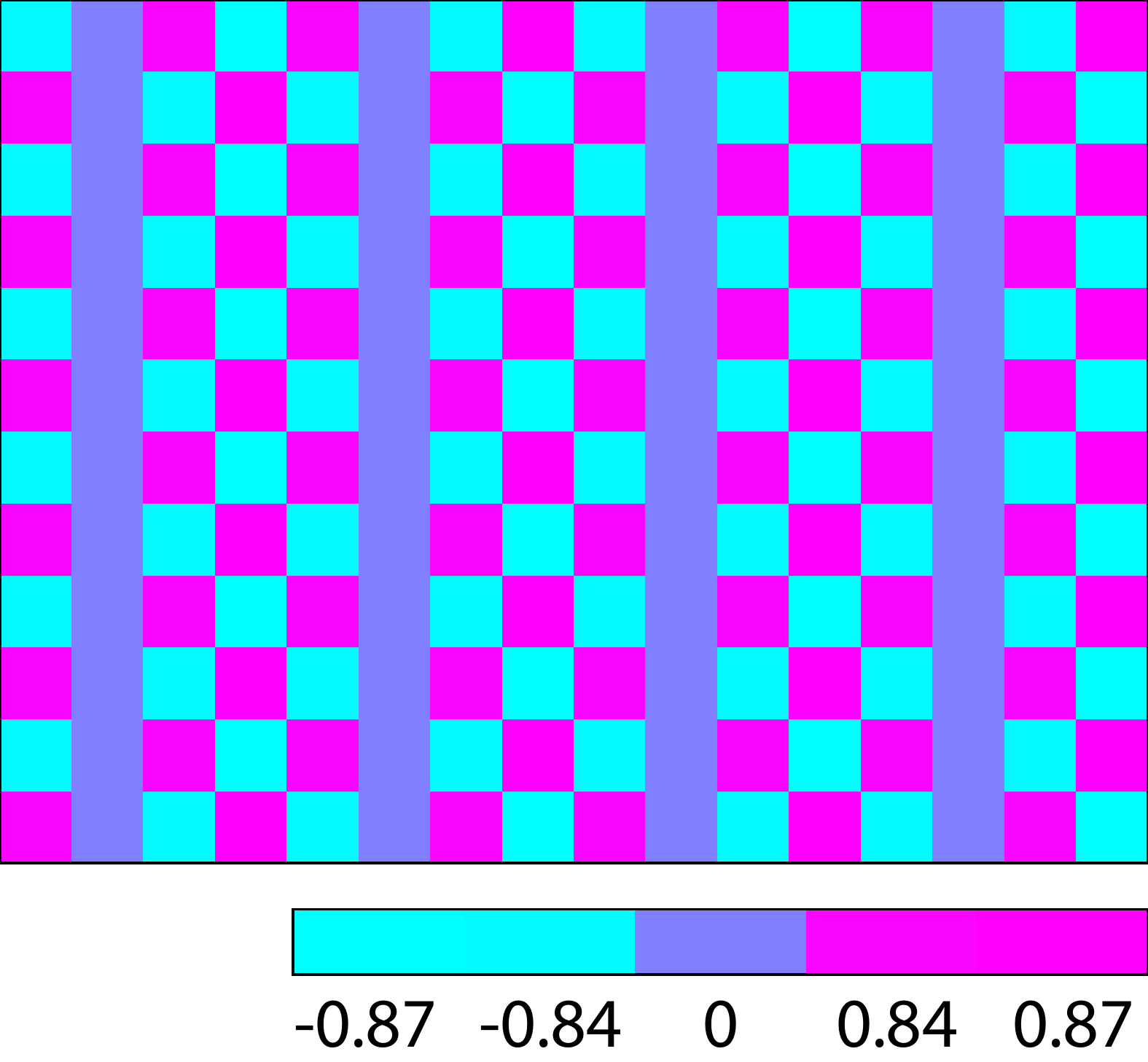}
\put(0,9){\bf(a)}
\end{overpic}\hspace{3mm}
\begin{overpic}
[width=0.64\columnwidth]{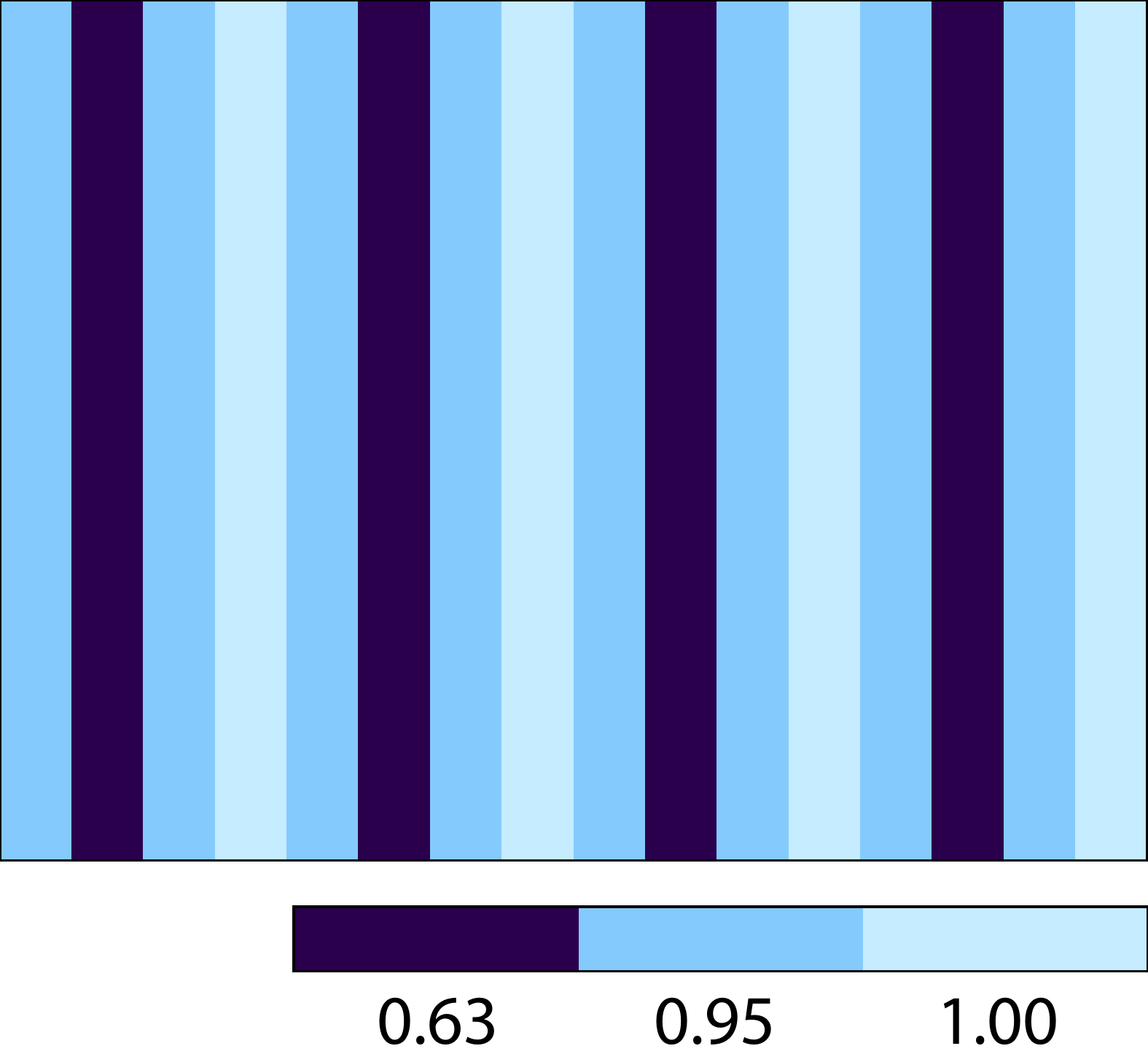}
\put(0,9){\bf(b)}
\end{overpic}\hspace{3mm}
\begin{overpic}
[width=0.64\columnwidth]{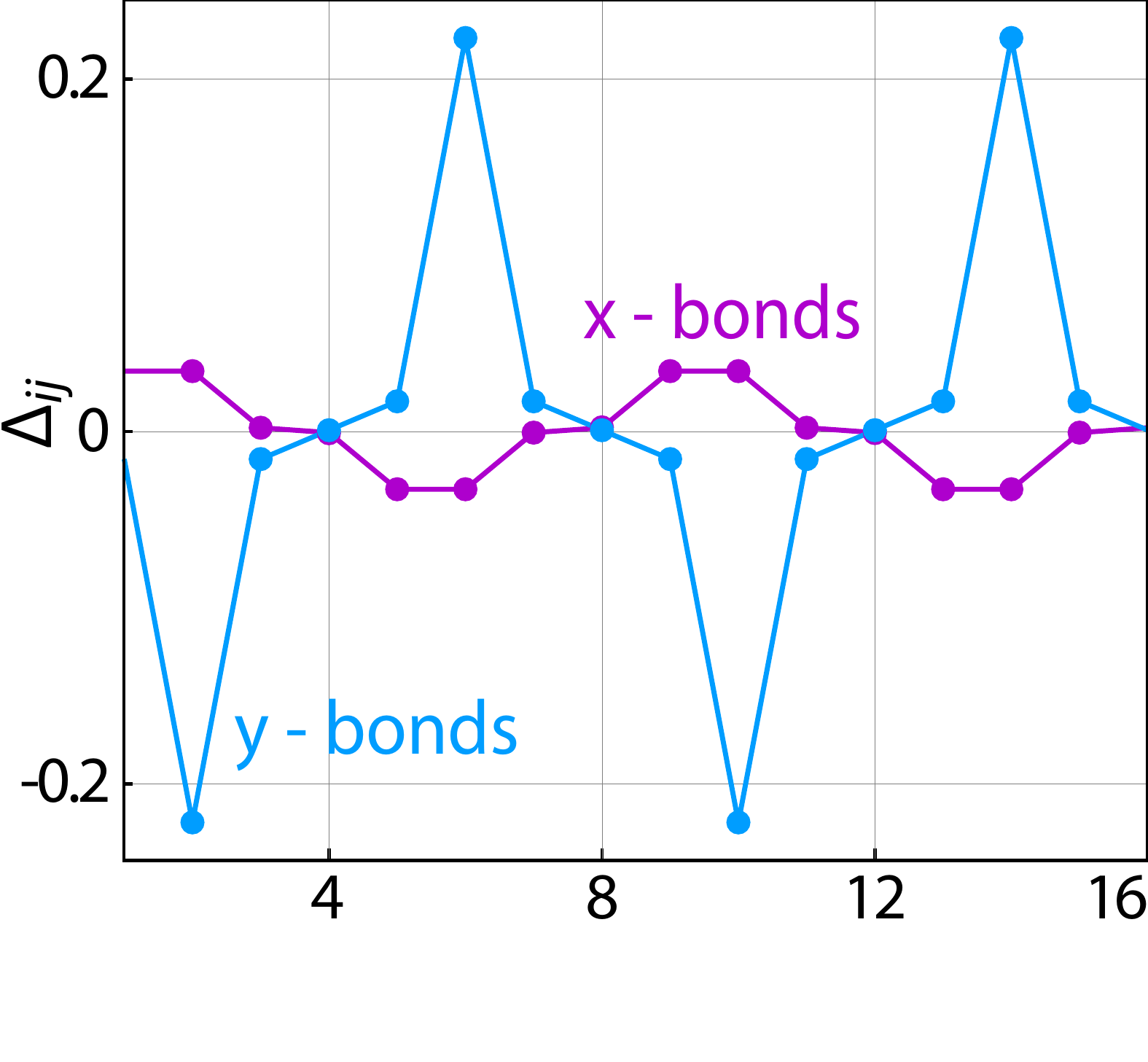}
\put(3,9){\bf(c)}
\end{overpic}
\caption{Real-space characterization of the striped SC state. {\bf (a)} The 
magnetization $m_i=n_{i\ua}-n_{i\da}$ exhibits AF stripes separated by 
non-magnetic anti-phase domain walls. {\bf (b)} The charge density 
$n_i=n_{i\ua}+n_{i\da}$ reaches nearly one electron per site inside the AF 
stripes which are separated by single lines near quarter filling, resulting in 
an overall mean charge density $7/8$, i.e. hole doping $1/8$. {\bf (c)} 
The SC bond order parameter on the horizontal (purple) and on the vertical bonds 
(blue). It is largest on the vertical bonds along the line of non-magnetic 
sites. Results were obtained for $V=2\,t$ on a 16$\times$12 lattice.
}
\label{Fig1}
\end{figure*}

Although striped SC states were encountered before 
\cite{Raczkowski07,Andersen05,Yang:2009}, a simple microscopic model 
Hamiltonian which supports a superconducting PDW groundstate with
spin and charge stripes has been lacking so far. Here we elaborate on the existence 
of these solutions in an isotropic 2D pairing Hamiltonian, characterize 
their real- and momentum-space properties, and relate them to existing experimental data. 

Our model is a tight-binding Hamiltonian ${\cal H}={\cal H}_0+{\cal H}_{\rm I}$,
where
$
{\cal H}_0=-\sum_{i,j}\sum_st_{ij}c^\dag_{is}c_{js}
\label{j0}
$
describes the hopping motion of free electrons on a square lattice. The 
operator $c_{js}$ ($c^\dagger_{js}$) annihilates (creates) an electron on 
lattice site $j$ with spin $s=\uparrow,\downarrow$; $t_{ij}$ are hopping matrix
elements with amplitude $t$ between nearest-neighbo, and $t'$ between 
next-nearest neighbor sites. Here we use $t'=-0.4\,t$ for all calculations. 
The BCS-type attractive interaction 
\begin{align}
{\cal H}_{\rm I}=-\frac{V}{2}\sum_{\langle i,j\rangle,s}c^\dag_{is}c^\dag_{j-s}
c_{j-s}c_{is}
\label{j1}
\end{align}
is restricted to nearest-neighbor sites; $V>0$ is the pairing interaction 
strength. In the complete mean-field decoupling scheme
\begin{align}
\begin{split}
{\cal H}_{\rm I}\longrightarrow\frac{1}{2}\sum_{\langle i,j\rangle}\Big[
\Delta^{\!*}_{ji}c_{j\da}c_{i\ua}+\Delta_{ij}c^\dag_{i\ua}c^\dag_{j\da}-Vn_{j\da}
c^\dag_{i\ua}c_{i\ua}\\
-Vn_{j\ua}c^\dag_{i\da}c_{i\da}+\frac{\displaystyle
\Delta^{\!*}_{ji}\Delta_{ij}}{V}+Vn_{i\ua}n_{j\da}\Big]
\end{split}
\label{j2}
\end{align}
we introduce the bond order parameter $\Delta_{ij}=-V\langle c_{j\da}c_{i\ua}
\rangle$ for superconductivity and the local spin resolved densities $n_{is}=\langle c^\dag_{is}c_{is}
\rangle$. Using a Bogoliubov-de Gennes transformation, the model is solved 
self-consistently at an electron density $7/8$ (for details on the formalism see e.g. Ref.~\onlinecite{Schmid}). The terms $Vn_{j,-s}c^\dag_{is}c_{is}$ are typically 
not accounted for in the standard BCS
theory. However, for a nearest-neighbor pairing interaction they are a strong 
source for antiferromagnetism, since they provide an energy gain $-V$ for 
each AF bond, but only $-V/4$ for a bond between two non 
spin-polarized sites. This is the driving force for the formation of 
AF stripes in our model.

There are two qualitatively different regimes of interaction strengths: for 
weak $V$, below a critical interaction strength $V_{{\rm c}1}\approx 0.9\,t$, the 
only solution of the self-consistency equations is a homogeneous SC 
phase with $d$-wave symmetry and without antiferromagnetism. As $V$ is increased beyond 
$V_{{\rm c}1}$, there is a sharp crossover into a regime where antiferromagnetism is the dominant 
order, superconductivity is suppressed and eventually disappears above a second critical 
interaction strength $V_{{\rm c}2}\approx3\,t$. The characteristics of this 
latter regime are best illustrated for  strong interactions $V>V_{{\rm c}2}$ 
when ${\cal H}_{\rm I}$ favors energetically a homogeneous 
AF phase for a half-filled band.  
For a filling $\rho = 7/8$ instead a configuration is preferred 
with three-legged, half-filled spin ladders and non-magnetic lines at an 
average density $\rho=1/2$ in between. 
This regularly striped solution which is unique for the filling $\rho=7/8$ was
indeed inferred from elastic neutron scattering data for the low-temperature 
phase of La$_{1.48}$Nd$_{0.4}$Sr$_{0.12}$CuO$_4$~\cite{TranquadaLNdSCO}.

If the interaction strength $V$ is reduced below $V_{{\rm c}2}$, 
 superconductivity emerges and resides predominantly on the quarter-filled 
channels in between the AF stripes. Figure~\ref{Fig1} displays the 
self-consistently determined magnetization $m_i=n_{i\ua}-n_{i\da}$ (a), the 
charge density $n_i=n_{i\ua}+n_{i\da}$ (b), and the SC order parameter (c)
of the striped SC state for $V=2\,t$ on a $16\times12$ lattice. 
The results presented here are stable groundstate solutions irrespective of the
system size, provided that the selected geometry is commensurate with the 
wavelength of the stripes. 
Along the AF stripes the magnetic energy gain is maximized by a nearly perfect 
antiparallel spin alignment.  
While stripe formation minimizes the magnetic energy, kinetic energy is gained
by transverse fluctuations. This is the origin of the sign change in the 
AF order between neighboring stripes. 

The unidirectional character of the SC order parameter is evident from its 
considerably smaller values on the bonds perpendicular to the stripe 
orientation (see Fig.~\ref{Fig1}c). The SC order parameter acquires its maximum 
value on the bonds which connect to the sites in the non-magnetic channels. 
The sign change between the SC order parameters on the 
horizontal and the vertical bonds connected to the same site verifies the 
$d$-wave character of the SC order parameter.

\begin{figure}[t!]
\centering
\vspace{2.5mm}
\begin{overpic}
[width=0.48\columnwidth]{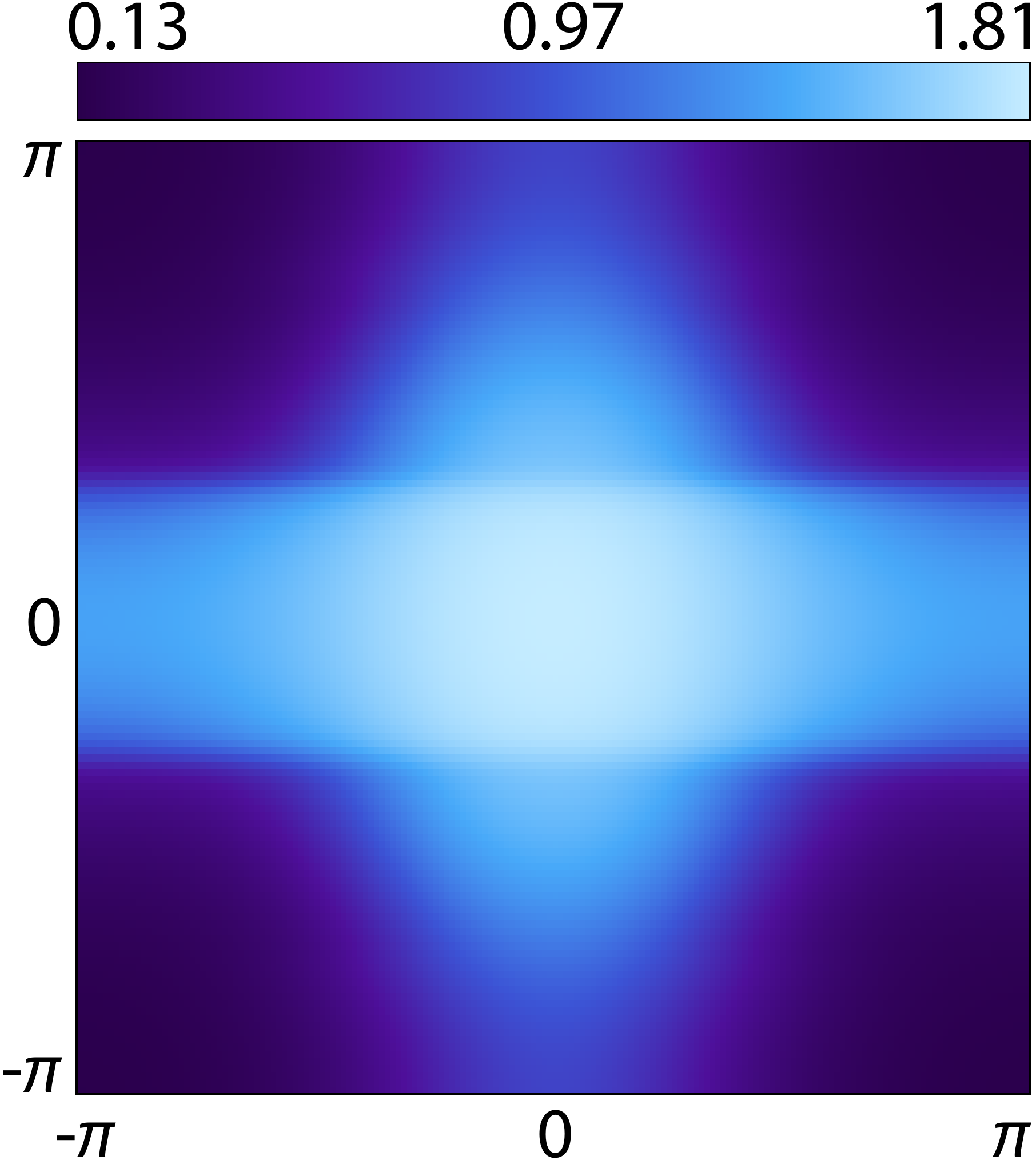}
\put(0,103){{\bf(a)}\ \ momentum distribution}
\end{overpic}\hspace{2mm}
\begin{overpic}
[width=0.48\columnwidth]{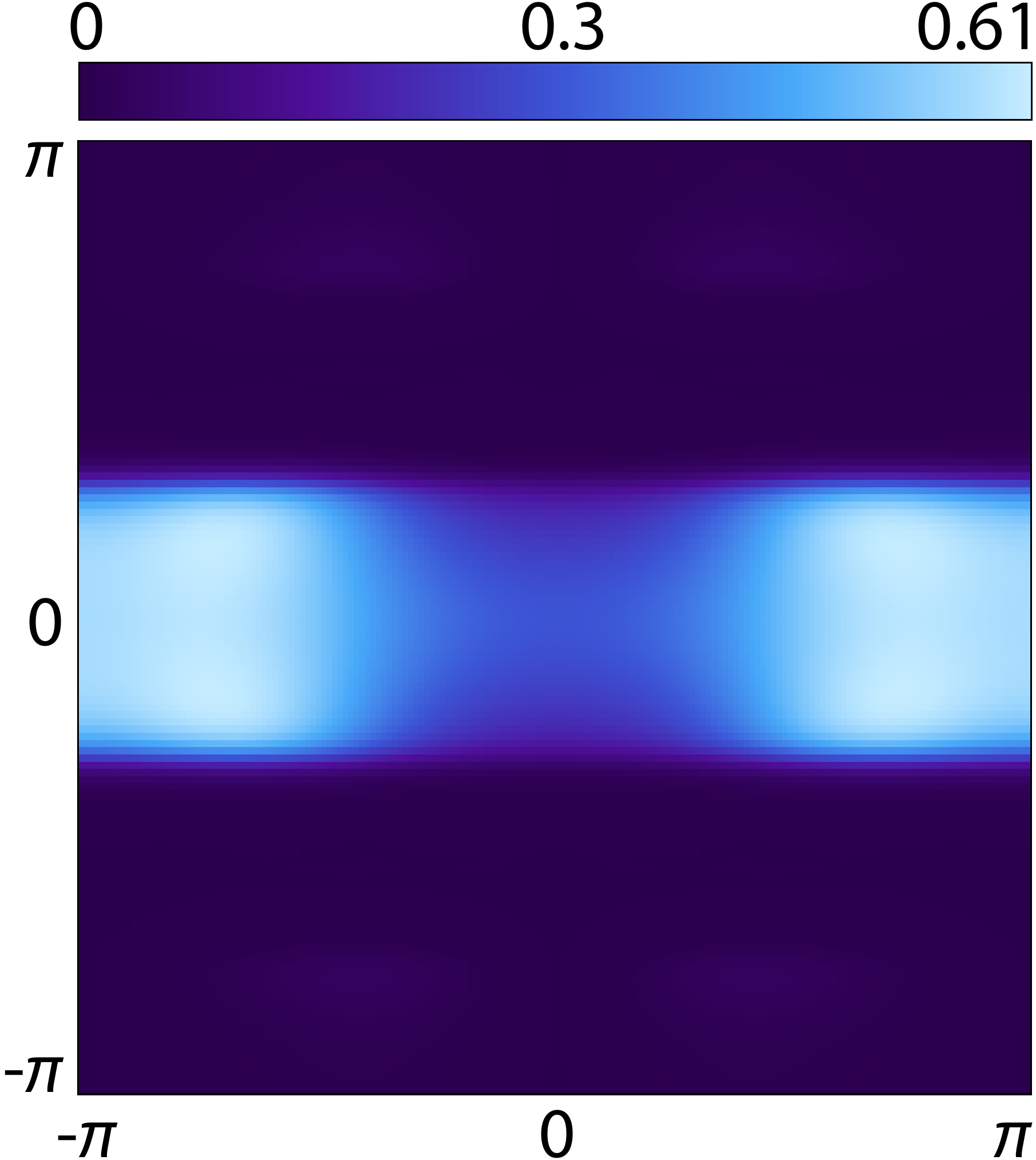}
\put(0,103){{\bf(b)}\ \ spectral density}
\end{overpic}\\\vspace{5mm}
\begin{overpic}
[width=0.48\columnwidth]{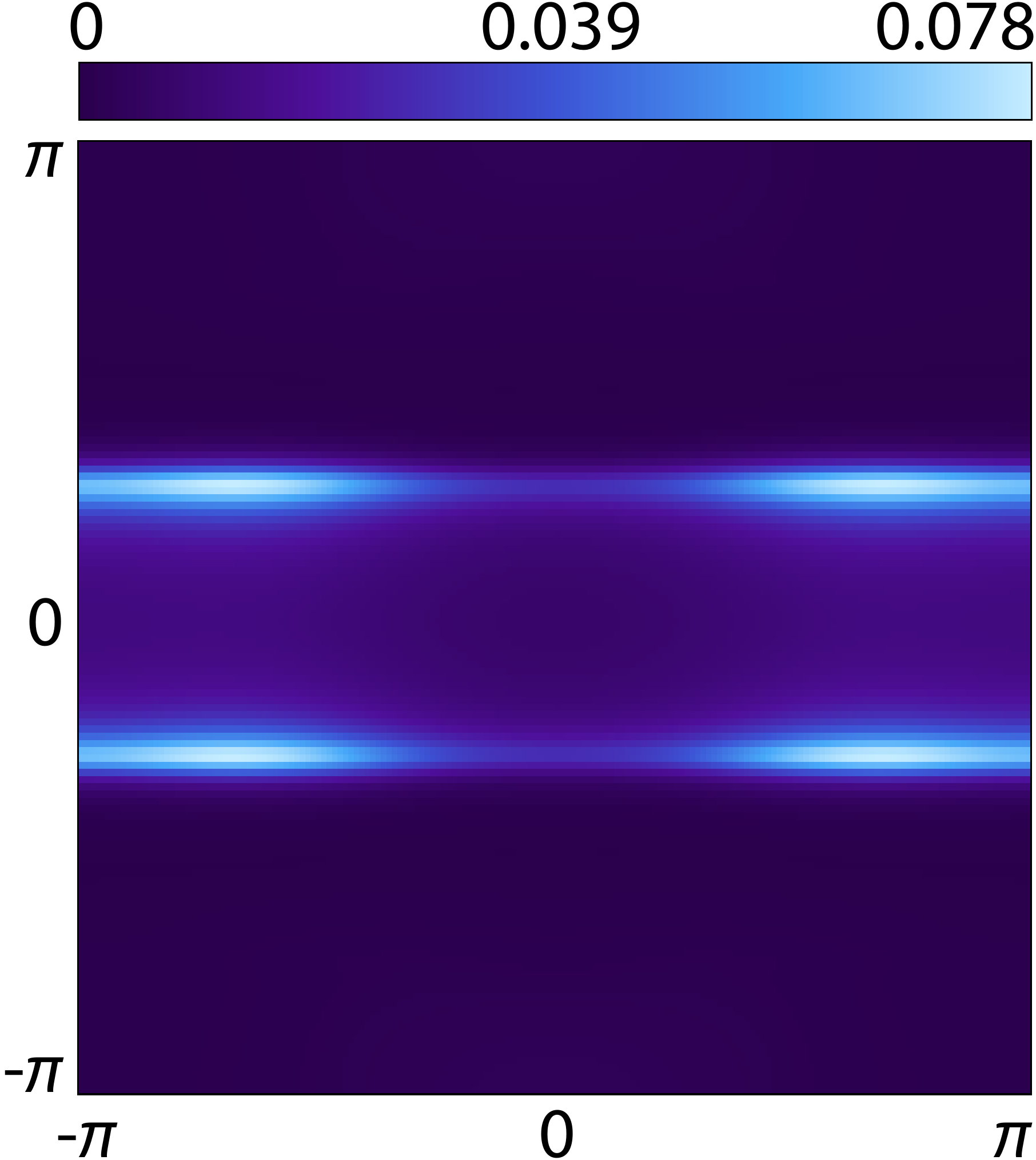}
\put(0,103){{\bf(c)} pair density}
\end{overpic}\hspace{2mm}
\begin{overpic}
[width=0.48\columnwidth]{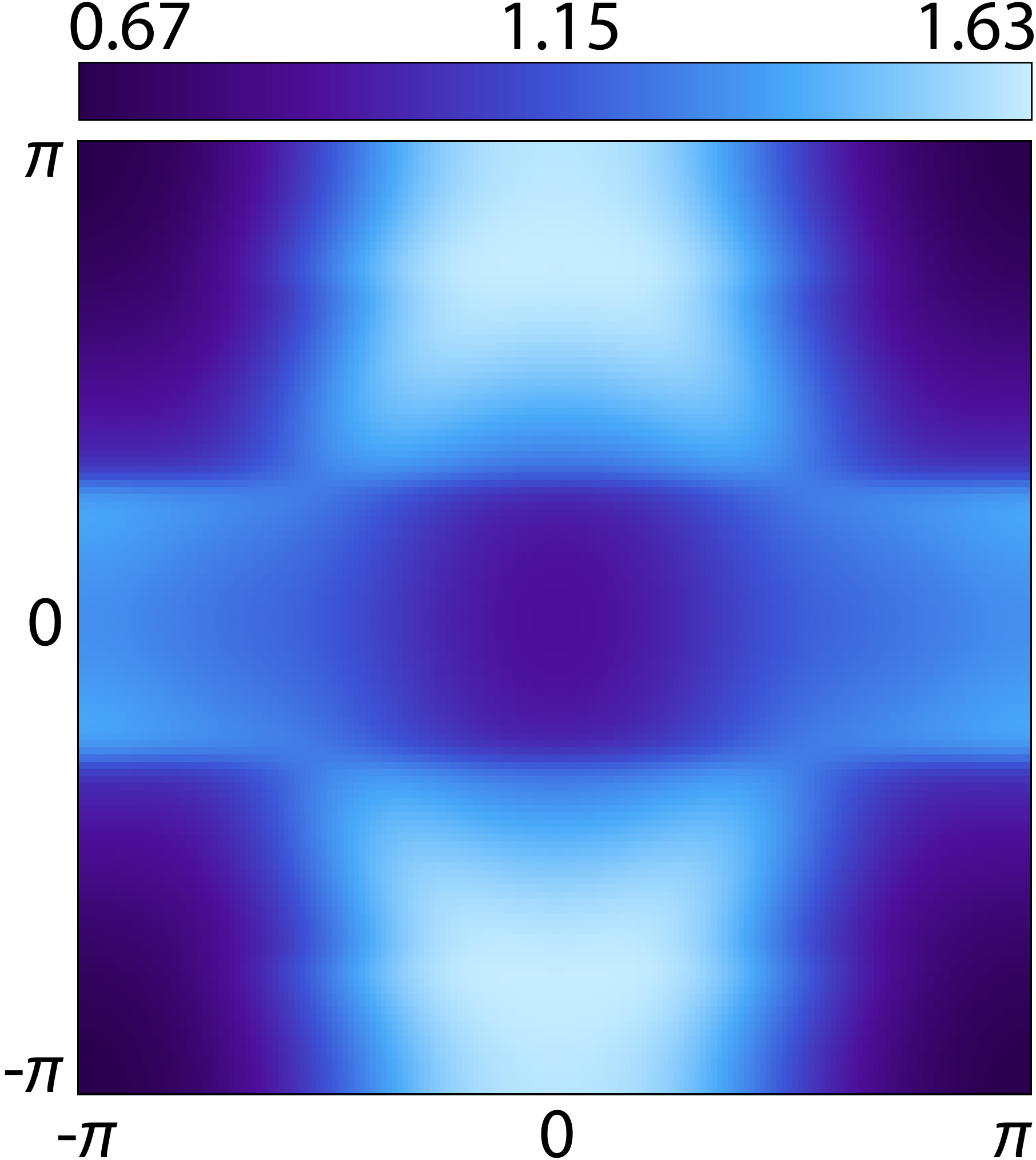}
\put(0,103){{\bf(d)}\ \ spin density}
\end{overpic}
\caption{Momentum-space characterization of the striped superconductor {\bf (a)}, 
The momentum distribution $n({\bf k})$ is a superposition of incoherent states
well below the Fermi energy, and a partially filled quasi one-dimensional band 
(horizontal bar). {\bf (b)}, The integrated spectral weight (see 
Eq.~(\ref{intsw})) captures only the conducting states originating from the 
motion along the stripes. {\bf (c)}, The pair density $P ({\bf k})$ shows the 
distribution of the SC pairs in momentum space. {\bf (d)}, The Fourier 
transformed spin density $\rho_S ({\bf k})$ is largest near $(0,\pm\pi)$ for
vertical stripes. Results were obtained for $V=2\,t$ on a 16$\bm\times$12 lattice
using 9$\bm\times$9 supercells.
}
\label{Fig2}
\end{figure}

The groundstate solution with striped superconductivity is not unique 
with respect to a sign change of $\Delta_{ij}$ between neighboring hole-rich 
channels. A solution degenerate to the one shown in Fig.~\ref{Fig1} exists 
without this sign change. Since in our model analysis all physical quantities 
depend on $\Delta^2_{ij}$ only, these two variants of the striped 
superconductor have the same energy, provided $\Delta_{ij}$ vanishes at the 
center of the AF stripes. For interaction strengths close to $V_{{\rm c}1}$, 
where the AF order weakens and $\Delta_{ij}$ becomes finite also within the AF 
stripes, this degeneracy is lifted and the state without the sign change
is favored. A similar conclusion was reached within a renormalized 
mean-field theory for a generalized $t$-$J$ model by Yang {\it et al}., if the 
hopping amplitudes are anisotropic~\cite{Yang:2009}. 

A qualitative difference of the two striped SC states with and without sign 
change of the order parameter concerns the center-of-mass momenta 
$\bf q$ of the electron pairs. In the former state all pairs have either 
momenta $q_x=\pm\pi/4a$ or $q_x=\pm3\pi/4a$ with $q_y=0$, corresponding to the 
periodicity of eight lattice sites. The state without sign change has in 
addition a finite ${\bf q}={\bf 0}$-component for pairs with vanishing total 
momentum, i.e. $\langle c_{{\bf k}\ua}c_{-{\bf k}\da}\rangle\neq0$. The finite 
center-of-mass momenta coexisting with ${\bf q}=\bm0$ are stabilized only by 
the AF stripes and vanish together with stripe order, thereby recovering the 
homogeneous $d$-wave superconductor. The state with sign change however remains
striped for sufficiently large $V$ even in the absence of AF order and realizes 
the pure PDW state discussed in Ref.~\onlinecite{Loder:2010}.

\begin{figure}[t!]
\centering
\includegraphics[width=0.98\columnwidth]{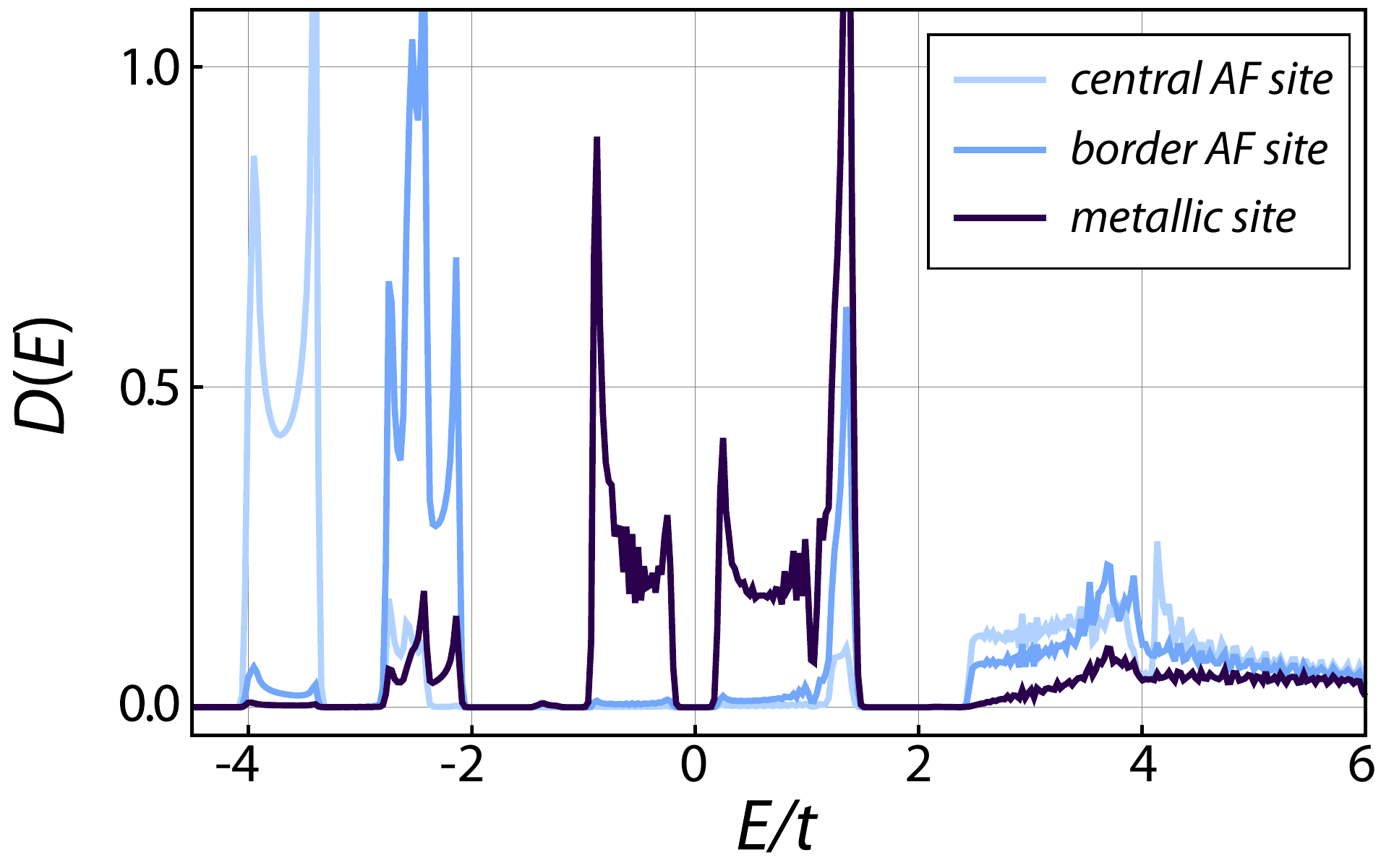}
\vspace{-1mm}
\caption{
{ Local density of states (LDOS).} The colors correspond to the three 
distinct sites of the striped superconductor: sites with minimum charge density
(dark blue), sites in the center (blue) and on the edge (light blue) of the AF stripes. 
} 
\label{Fig3}
\end{figure}

In Fig.~\ref{Fig2} the striped superconductor is characterized in momentum 
space. The calculations were performed on a $16\times12$ lattice with 
$9\times9$ supercells to ensure satisfactory momentum resolution. The momentum 
distribution $n({\bf k})$ in Fig.~\ref{Fig2}a clearly exhibits the 
unidirectional character of the striped system with a horizontal bar of high 
occupation probability and a diffuse region around the Brillouin zone (BZ) 
center. This diffuse background traces the original 2D Fermi surface of the 
uncorrelated electrons. In the absence of superconductivity the stripe order 
leads to a sharp Fermi surface with occupied states for momenta ${\bf k}$ with 
$k_y \leq \pi/4$. The absence of discontinuites in $n({\bf k})$ in 
Fig.~\ref{Fig2}a is due to a finite energy gap in the density of states (see 
Fig.~\ref{Fig3}). A remarkably similar momentum distribution has indeed 
been measured by Zhou {\it et al.} for the rare-earth doped cuprate 
La$_{1.28}$Nd$_{0.6}$Sr$_{0.12}$CuO$_4$ with static stripe order but no 
superconductivity~\cite{Zhou:1999}. For a comparison with the measured spectral
 weight we display the integrated spectral function 
\begin{align}
\int_{\mu-\omega_{\rm c}}^\mu A({\bf k},\omega)\,{\rm d}\omega = -\frac{1}{\pi}
\int_{\mu-\omega_{\rm c}}^\mu \mathrm{Im}\, G ({\bf k},{\bf k},\omega)\,{\rm d}\omega
\label{intsw}
\end{align}
in Fig.~\ref{Fig2}b. The lower energy cut-off at $\omega_{\rm c}=-t$ restricts the 
spectral weight to the contributions of the non-magnetic channels (c.f. 
Fig.~\ref{Fig3}). The chemical potential $\mu$ corresponds to 7/8 filling.
The quasiparticle excitations near the Fermi level occupy the
horizontal bar in momentum space with a strongly reduced spectral weight in the
center of the BZ as observed in the measurements of 
Ref.~\onlinecite{Zhou:1999}. This \textquotedblleft breach\textquotedblright\ 
in the spectral function is not captured by the physics of isolated spin 
ladders and indicates that the conducting states are not decoupled from the 
magnetic stripes. 

Similarly to Ref.~\onlinecite{Baruch:2008} we define the electron-pair density 
$P({\bf k})$ for singlet pairing as
$P^2 ({\bf k}) = \sum_{\bf q} 
\left\langle c_{-{\bf k}+{\bf q} \downarrow} c_{{\bf k}\uparrow} \right\rangle^2$.
$P({\bf k})$ serves as a measure at which momenta in the BZ electron pairs 
predominantly form in the superconductor. For the striped superconductor 
discussed above electron pairs have the finite center of mass momenta 
$\pm{\bf q}$ where $q_y=0$ and $q_x=\pi/4$ or $3\pi/4$ 
according to a stripe wavelength of 8 lattice constants.
The pair density is expected to be largest near the Fermi surface of the normal
conducting system as is indeed verified in Fig.~\ref{Fig2}c.
In a similar way we also translate the spin-stripe pattern into momentum 
space
$
 \rho_S ({\bf k})=\sum_{\bf q}\sum_s\langle s
c_{{\bf k}+{\bf q}s}^\dagger c_{{\bf k}s}\rangle.
$
As shown in Fig.~\ref{Fig2}d, $\rho_S({\bf k})$ is strongest near $(0,\pm\pi)$ 
for vertically oriented AF stripes, i.e., in those regions of the BZ where no 
electron pairs form.

\begin{figure}[t!]
\centering
\vspace{3mm}
\includegraphics[width=0.98\columnwidth]{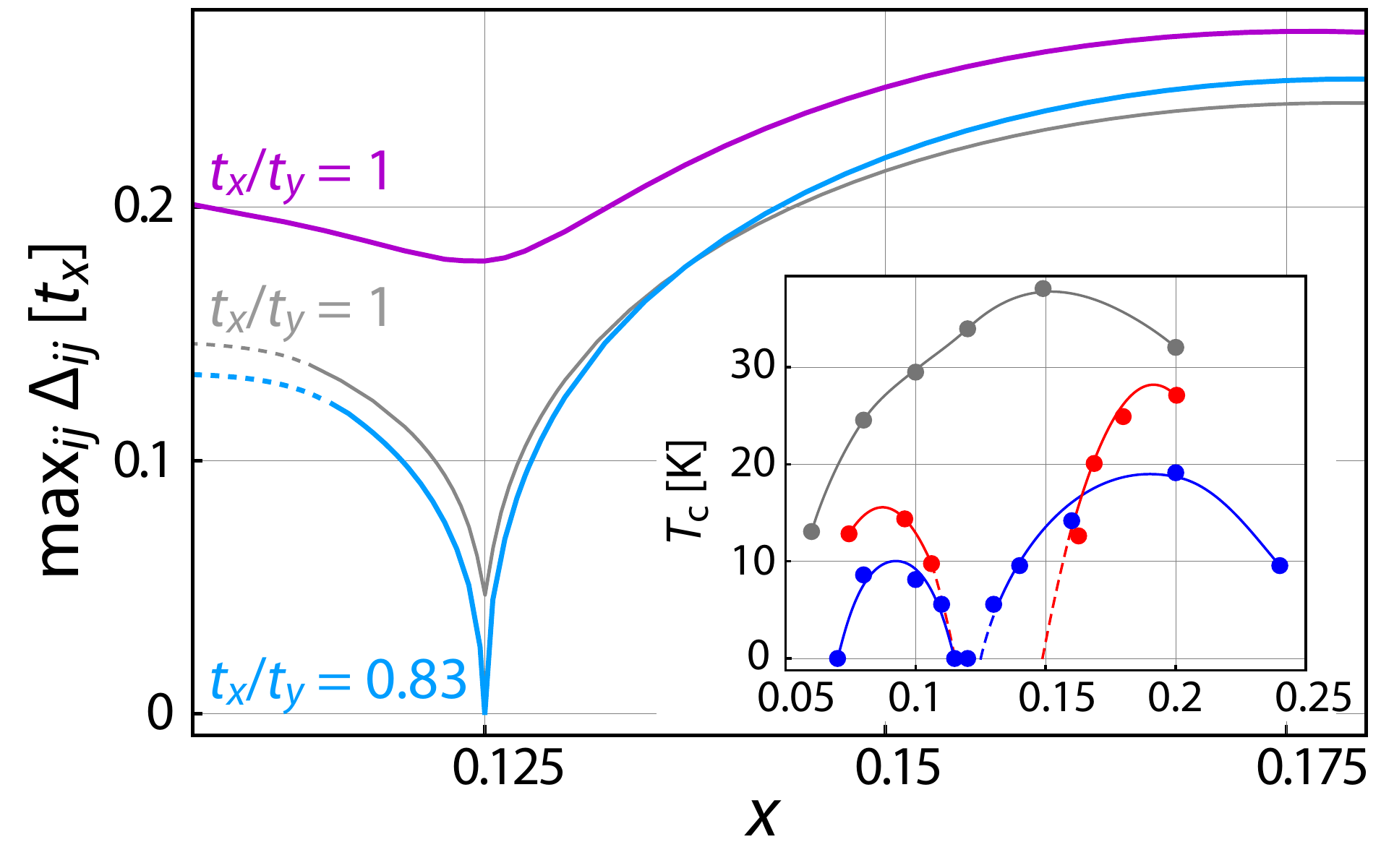}
\vspace{-3mm}
\caption{
{ Doping dependence of the superconducting order parameter.} The purple and 
blue curves for the maximum SC order parameter $\max_{ij}\Delta_{ij}$ 
correspond to $V=2t$ for isotropic $t_x=t_y$ and anisotropic hopping 
$t_x/t_y=0.83$, respectively. The grey line is the result for $t_x=t_y$ with a 
reduced $V=1.8t$. The dashed lines indicate an extrapolation to values of $x=1$
where the numerical procedure does not converge to the discussed solution. 
Inset: Measured doping dependence of $T_{\rm c}$ in 
La$_{2-y-x}$Nd$_{y}$Sr$_x$CuO$_4$ for $y=0.2$ (Ref.~\onlinecite{Buchner94}) 
(red), $y=0.4$ (dark blue), and $y=0$ (grey) (Ref.~\onlinecite{Tranquada97}). 
}
\label{Fig4}
\end{figure}

The absence of a discontinuity in $n({\bf k})$ is tied to the opening of a full
gap in the DOS. The local DOS in Fig.~\ref{Fig3} shows a large energy gap 
in the center of the AF stripes which is reduced on the edge of the stripe. But
also on the non-magnetic sites, where the SC order parameter is strongest, 
there exists a small gap, because the quasi one-dimensionality admixes a 
significant extended $s$-wave component. This is in contrast to the rather 2D 
pure PDW state without antiferromagnetism, where the local DOS is 
gapless.~\cite{Baruch:2008,Berg:2009,Loder:2010} 

The special affinity to stripe formation at the density $\rho=7/8$ in the 
cuprates is evident from the variation of the superconducting transition 
temperature $T_{\rm c}$. The striped compounds La$_{2-y-x}$Nd$_{y}$Sr$_x$CuO$_4$
\cite{Buchner94,Tranquada97} and La$_{15/8}$Ba$_{1/8}$CuO$_4$ 
\cite{Tranquada08,Hucker10} show a sharp dip in $T_{\rm c}(x)$ for hole doping 
$x=1-\rho=1/8$. The observed reduction of $T_{\rm c}$ is even stronger when 
lattice anisotropies in the low-temperature tetragonal (LTT) phase grow with 
increasing Nd content (c.f. insert in Fig.~\ref{Fig4}). Our model calculations 
reproduce these features as is evident from the doping dependence of the 
maximum SC order parameter $\Delta=\max_{ij}\Delta_{ij}$ shown in 
Fig.~\ref{Fig4}. 
If we simulate the lattice anisotropy from the octahedral tilt in the LTT phase
by introducing an anisotropy in the hopping amplitudes $t_x\neq t_y$, 
superconductivity is weakened and the minimum at $x=1/8$ develops into a sharp 
dip, which reaches $\Delta=0$ for $t_x/t_y\lesssim0.83$. For weaker pairing 
interaction strengths the dip at $x=1/8$ develops also for isotropic hopping 
$t_x=t_y$. When $x$ is decreased, the electron density in the SC stripes 
increases, and sequentially at specific values of $x$ the conducting stripes 
 turn antiferromagnetic one by one. This process is indicated by the dashed 
lines in Fig.~\ref{Fig4}. Since $\Delta$ is a measure of $T_{\rm c}$, the 
results for $\Delta$ can directly be compared to the $T_{\rm c}$ data for 
La$_{2-y-x}$Nd$_{y}$Sr$_x$CuO$_4$~\cite{Buchner94,Tranquada97} in the inset of 
Fig.~\ref{Fig4}. There is almost no suppression of $T_{\rm c}$ for the isotropic
compound with $y=0$, but a complete destruction of superconductivity around 
$x=1/8$ in the anisotropic Nd-doped compounds with $y=0.2$ and $y=0.4$.

The presented pairing model for the coexistence of SC and AF stripe order 
reproduces the most prominent properties of striped high-$T_{\rm c}$ cuprates 
remarkably well. Although it was shown before that the pure PDW can be the 
groundstate of a pairing Hamiltonian in the absence of 
magnetism~\cite{Loder:2010}, the issue concerning the sign change of the SC 
order parameter will only be resolved by a phase sensitive extension of the 
present calculation, e.g. a Josephson coupling term which has to be identified
beyond the Hartree-Fock decoupling scheme~\cite{Berg09}.
Defects may certainly affect the stability of the stripe state considerably;
the pure PDW is indeed supposed to be fragile with respect to
impurities.~\cite{Berg09} As we have verified, the inclusion of potential scatterers in our model 
shows that superconductivity is weakened and eventually vanishes. 
The AF stripe order, however, is affected only little. Moreover, impurities can
act as pinning forces for fluctuating stripes and thereby support the 
formation of static stripe order.

The authors gratefully acknowledge discussions with 
Brian M. Andersen, Raymond Fr\'esard, Peter Hirschfeld, and Steve Kivelson. 
This work was supported by the DFG (TRR 80).

\vspace{5mm}

\end{document}